\documentclass[sigconf]{acmart}
\def\BibTeX{{\rm B\kern-.05em{\sc i\kern-.025em b}\kern-.08emT\kern-.1667em\lower.7ex\hbox{E}\kern-.125emX}}
    
\copyrightyear{2019} 
\acmYear{2019} 
\setcopyright{acmlicensed}
\acmConference[DaMoN'19]{International Workshop on Data Management on New Hardware}{July 1, 2019}{Amsterdam, Netherlands}
\acmBooktitle{International Workshop on Data Management on New Hardware (DaMoN'19), July 1, 2019, Amsterdam, Netherlands}
\acmPrice{15.00}
\acmDOI{10.1145/3329785.3329931}
\acmISBN{978-1-4503-6801-8/19/07}

\begin{document}

%
% The "title" command has an optional parameter, allowing the author to define a "short title" to be used in page headers.

\title{Persistent Buffer Management with Optimistic Consistency}

%
% The "author" command and its associated commands are used to define the authors and their affiliations.
% Of note is the shared affiliation of the first two authors, and the "authornote" and "authornotemark" commands
% used to denote shared contribution to the research.
\author{Lucas Lersch}
\email{lucas.lersch@sap.com}
\affiliation{%
  \institution{TU Dresden \& SAP SE}
}
\author{Wolfgang Lehner}
\email{wolfgang.lehner@tu-dresden.de}
\affiliation{%
  \institution{TU Dresden}
}
\author{Ismail Oukid}
\email{ismail.oukid@sap.com}
\affiliation{%
  \institution{SAP SE}
}

%
% By default, the full list of authors will be used in the page headers. Often, this list is too long, and will overlap
% other information printed in the page headers. This command allows the author to define a more concise list
% of authors' names for this purpose.
\renewcommand{\shortauthors}{Lersch et al.}

%
% The abstract is a short summary of the work to be presented in the article.
\begin{abstract}
Finding the best way to leverage non-volatile memory (NVM) on modern database systems is still an open problem.
The answer is far from trivial since the clear boundary between memory and storage present in most systems seems to be incompatible with the intrinsic memory-storage duality of NVM.
Rather than treating NVM either solely as memory or solely as storage, in this work we propose how NVM can be simultaneously used as both in the context of modern database systems.
We design a persistent buffer pool on NVM, enabling pages to be directly read/written by the CPU (like memory) while recovering corrupted pages after a failure (like storage).
The main benefits of our approach are an easy integration in the existing database architectures, reduced costs (by replacing DRAM with NVM), and faster peak-performance recovery.
\end{abstract}

%
% This command processes the author and affiliation and title information and builds
% the first part of the formatted document.
\maketitle

\section{Introduction}
NVM is a persistent media promising higher bandwidth (2$\times$) and lower latency (10$\times$) than modern NAND-flash SSDs.
Furthermore, NVM can be attached to the memory bus, thus allowing it to be directly accessed by the CPU through its caches in a much smaller granularity (cache-lines) than regular block devices.
Therefore, NVM introduces not only a new layer within the storage hierarchy~\cite{bonnet2017s}, but it also enables more flexibility regarding data placement.

While a few factors slower, reading data directly from NVM can be done the same way as with DRAM.
However, writing data directly to NVM imposes challenges in terms of data consistency.
The root of these challenges is the lack of control the application has over data movement between CPU cache and NVM in comparison to the data movement between DRAM and SSD.
In other words, the programmer cannot prevent cache lines from being evicted from the CPU cache and written-back to NVM at arbitrary points in time.

Related work address these challenges with solutions that fall in one of three categories (also identified by previous work~\cite{van2018managing}).
First, early proposals leverage NVM as a cheaper alternative to extend DRAM, enabling larger buffer pools~\cite{wu2014app,ou2014wear}.
These approaches focus on reducing write amplification and improving wear leveling on NVM, but they do not enforce any consistency when writing data, and therefore do not leverage persistency.
Second, persistent data structures~\cite{oukid2016fptree,yang2015nv,chen2015persistent, arulraj2018bztree,lee2017wort} enable direct fine-grained writes to NVM by issuing out-of-place writes and relying on instruction ordering (\emph{SFENCE}) and eagerly flushing cache lines (\emph{CLFLUSH/CLWB}) to make the operation visible (usually by flipping a validity bit).
Third, buffer managers were proposed to integrate NVM with the storage hierarchy below DRAM and above SSD~\cite{pelley2013storage,van2018managing,eisenman2018reducing}.
These approaches access data on DRAM, and therefore they have full control of when data is persisted.
Optimizations can be made regarding block sizes, but they still impose movement of data between DRAM and NVM.

The first two categories treat NVM more like traditional memory, while the last one treats NVM more like traditional storage.
However, NVM is actually both.
We consider that a database system should allow NVM to be read and written directly, thus exploring its memory-storage duality to its full potential.
Our system accesses NVM like memory, but it guarantees consistency of writes to NVM like storage.
We achieve that by integrating NVM in the buffer pool of a database system to either extend DRAM or completely replace it, while still leveraging its persistency in an optimistic way.
In other words, we never enforce ordering of writes or eagerly flush cache lines.
The main motivation is that corruption occurs when a write operation is partially evicted from the CPU cache to NVM.
With the capacity of NVM being significantly higher than CPU caches, the probability of corruption tends to be low and therefore pessimistically enforcing consistency of every write introduces a relatively high overhead by not leveraging the CPU cache.

Recent work~\cite{DBLP:journals/corr/abs-1901-10938} has gone into the same direction of allowing data to be accessed directly on NVM.
This approach complements our approach by focusing on cost models for optimizing data movement, while we focus on enabling direct writes to NVM to be consistent.

\section{Background}
We give an overview of the techniques that we use to implement a persistent buffer management in a database system.

\textbf{Buffer Management:}
We assume a traditional transactional storage manager having a B+Tree as its workhorse data structure.
The B+tree is organized such that nodes are represented by pages, which are the unit of data movement and buffering.
Optimizations like pointer swizzling~\cite{graefe2014memory} and low-overhead replacement policies~\cite{leis2018leanstore} may apply.
The atomicity and durability of writes to pages buffered in DRAM is guaranteed by write-ahead logging (WAL).

\textbf{Single-page Recovery:}
Similar to traditional ARIES~\cite{mohan1992aries}, we assume a page-level physiological logging.
This implies that pages are not only a unit of data movement, but also of fault containment and repair~\cite{DBLP:journals/pvldb/GraefeK12}.
This enables techniques such as write-elision, on-demand instant restart and restore, and single-page repair~\cite{graefe2014instant}.
These techniques are the base to enable direct writes directly made to NVM to be consistent without eagerly flushing cache lines.

\section{System Design}
\begin{figure}
    \centering
    \includegraphics{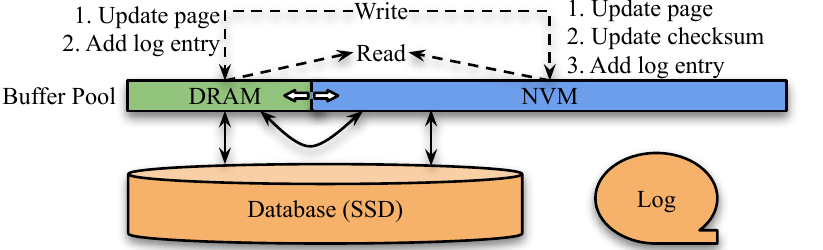}
    \caption{During normal processing reads and writes can access both DRAM and NVM directly (dashed lines). Whole pages can be moved across any devices (solid lines). The trade-offs between DRAM and NVM can be analysed by sliding an abstract persistency bar.}
    \label{fig:nvm_buffer}
\end{figure}

The pages of our system are primarily located on SSD and only the warm pages are buffered in NVM.
Hot pages might be buffered in a DRAM portion of the buffer pool, as seen in Figure~\ref{fig:nvm_buffer}.

\textbf{Normal Processing: }
During normal transactional processing, a page to be updated will be either on SSD, NVM, or DRAM.
In case the page is on SSD, it is loaded to the buffer pool (either to DRAM or NVM, to be decided by a placement policy).
In case the page is on DRAM, we have a hit and the page is updated normally.
If the page is on NVM, two actions might occur.
First, the page might be identified as "heating up" by a placement policy such as 2Q~\cite{lersch2017rethinking}, in which case it will be copied to DRAM and updated there.
Second, the page might be simply warm, in which case the update is done directly on NVM.
In the last case, since atomicity and durability is guaranteed by WAL, issuing \emph{CLFLUSH/CLWB} after updating the page is not necessary as the log serves as the single source of truth.
However, when restarting after a system failure, pages that were on  NVM might be corrupted because updates were not properly persisted.
As a consequence, the current state of a page on NVM is unknown and therefore persistency of NVM cannot be leveraged.

We address this challenge by dividing it in two dimensions: corruption detection and page repair.
Each page contains an 8 Byte checksum of the whole page.
Whenever the content of a page is modified directly on NVM, the page checksum is re-calculated and updated.
At this point, the modifications and the checksum of a page might be persistent or not, since we do not explicitly flush them from the CPU cache to NVM.
In case the overhead of calculating the checksum for the whole page after every update is too high, it can be reduced either by updating the checksum only after a certain number of updates (at the cost of higher corruption ratio) or by introducing multiple checksums per page corresponding to fractions of the page (at the cost of higher space consumption).

\textbf{Corruption Detection}
After a system failure, the restart process starts with log analysis, which identifies the state (not the content) of pages right before the crash.
We assume state-of-the-art page-based on-demand restart~\cite{graefe2014instant}, therefore a page being requested might still be lingering in the NVM portion of the buffer pool since before the crash.
Two steps are required to determine if the page can be used.
First, its checksum is calculated and compared to the checksum stored within the page.
If the checksums do not match, it is because either the checksum or the updated data were not evicted from the CPU cache to NVM.
Second, if the checksums match, the log sequence number of the last update in the page (\emph{pageLSN}) is compared to the LSN retrieved during log analysis (\emph{expectedLSN}).
To summarize, a page might be in one of the following states:

\begin{itemize}
    \item Corrupted, if checksum does not match
    \item Behind, if checksum matches and $ pageLSN < expectedLSN $
    \item Current, if checksum matches and $ pageLSN = expectedLSN $
    \item Ahead, if checksum matches and $ pageLSN > expectedLSN $
\end{itemize}

The page is \emph{behind}, if it was updated by a committed transaction (the log records were flushed) but neither the update nor the checksum were evicted from the CPU cache.
In this case, the page is in a consistent but outdated state since it violated the write-ahead rule.
However, this violation is tolerated on NVM as long as it is guaranteed not to happen on SSD.
Analogously, the page is \emph{ahead} if both update and checksum were persisted to NVM but the transactions that made these updates did not commit.
Finally, if both checksum and LSNs match, the page is in its most \emph{current} state and ready to be accessed.
Except for the \emph{current} state, all other states must be recovered by replaying log records.

\textbf{Page Repair}
No assumption can be made about a \emph{corrupted} page, and therefore the only alternative is to retrieve its more recent version from SSD (which acts as a backup) and replay the log records referring to this page to bring it up to its most recent state.
A \emph{behind} page is missing committed updates, therefore it can be used as the basis for replaying the log records, not requiring an older version of the page to be fetched from SSD.
\emph{Ahead} pages are consistent but contain updates made by uncommitted transactions.
Since ARIES-style recovery only allows logical UNDO on a transaction level, the updates of a single page cannot be independently rolled back.
Therefore, the same procedure of a \emph{corrupted} page applies for rolling forward.
Since it is required that an older version of the page is read from the database, it is convenient that frequently modified pages are flushed to the database often to bound the recovery time in such cases.
Fortunately, this can be achieved by an asynchronous cleaning job that iterates through the buffer pool and picks dirty pages to be flushed to the database.

\section{Conclusion}
In this work we propose extending database systems with a persistent buffer pool on NVM.
We consider such extension to impose a low implementation effort, since NVM is treated very similarly to DRAM during runtime, while leveraging well-understood recovery algorithms to enforce its consistency.
The consequence is that the persistency aspect can be leveraged in an optimistic way, without major changes in the code base to enforce ordering of writes to NVM.
Furthermore, cache lines do not have to be eagerly flushed, which enables write operations to be cached in the CPU and amortize the higher latencies of NVM.
We also consider that in the short and medium term, NVM will not replace neither DRAM or SSD, but should act in synergy with both.
Our design also enables the user to explore different proportions of DRAM and NVM in the buffer pool: more DRAM will lead to higher performance, while more NVM will lead to reduced costs and faster recovery (higher availability).
This adds more flexibility for analyzing trade-offs and navigating the performance continuum between modern expensive in-memory databases and more traditional low-cost disk-based systems.

%
% The next two lines define the bibliography style to be used, and the bibliography file.
\bibliographystyle{ACM-Reference-Format}
\bibliography{sample-base}


\begin{thebibliography}{18}

%%% the Web address.

\ifx \showCODEN    \undefined \def \showCODEN     #1{\unskip}     \fi
\ifx \showDOI      \undefined \def \showDOI       #1{#1}\fi
\ifx \showISBNx    \undefined \def \showISBNx     #1{\unskip}     \fi
\ifx \showISBNxiii \undefined \def \showISBNxiii  #1{\unskip}     \fi
\ifx \showISSN     \undefined \def \showISSN      #1{\unskip}     \fi
\ifx \showLCCN     \undefined \def \showLCCN      #1{\unskip}     \fi
\ifx \shownote     \undefined \def \shownote      #1{#1}          \fi
\ifx \showarticletitle \undefined \def \showarticletitle #1{#1}   \fi
\ifx \showURL      \undefined \def \showURL       {\relax}        \fi
% invisible to TeX
\providecommand\bibfield[2]{#2}
\providecommand\bibinfo[2]{#2}
\providecommand\natexlab[1]{#1}
\providecommand\showeprint[2][]{arXiv:#2}

\bibitem[\protect\citeauthoryear{Arulraj, Levandoski, Minhas, and
  Larson}{Arulraj et~al\mbox{.}}{2018}]%
        {arulraj2018bztree}
\bibfield{author}{\bibinfo{person}{Joy Arulraj}, \bibinfo{person}{Justin
  Levandoski}, \bibinfo{person}{Umar~Farooq Minhas}, {and}
  \bibinfo{person}{Per-Ake Larson}.} \bibinfo{year}{2018}\natexlab{}.
\newblock \showarticletitle{{BzTree: A High-Performance Latch-free Range Index
  for Non-Volatile Memory}}.
\newblock \bibinfo{journal}{\emph{{PVLDB}}} \bibinfo{volume}{11},
  \bibinfo{number}{5} (\bibinfo{year}{2018}), \bibinfo{pages}{553--565}.
\newblock


\bibitem[\protect\citeauthoryear{Arulraj, Pavlo, and Malladi}{Arulraj
  et~al\mbox{.}}{2019}]%
        {DBLP:journals/corr/abs-1901-10938}
\bibfield{author}{\bibinfo{person}{Joy Arulraj}, \bibinfo{person}{Andy Pavlo},
  {and} \bibinfo{person}{Krishna~Teja Malladi}.}
  \bibinfo{year}{2019}\natexlab{}.
\newblock \showarticletitle{{Multi-Tier Buffer Management and Storage System
  Design for Non-Volatile Memory}}.
\newblock \bibinfo{journal}{\emph{CoRR}}  \bibinfo{volume}{abs/1901.10938}
  (\bibinfo{year}{2019}).
\newblock
\showeprint[arxiv]{1901.10938}
\urldef\tempurl%
\url{http://arxiv.org/abs/1901.10938}
\showURL{%
\tempurl}


\bibitem[\protect\citeauthoryear{Bonnet}{Bonnet}{2017}]%
        {bonnet2017s}
\bibfield{author}{\bibinfo{person}{Philippe Bonnet}.}
  \bibinfo{year}{2017}\natexlab{}.
\newblock \showarticletitle{What's up with the storage hierarchy?}. In
  \bibinfo{booktitle}{\emph{CIDR}}.
\newblock


\bibitem[\protect\citeauthoryear{Chen and Jin}{Chen and Jin}{2015}]%
        {chen2015persistent}
\bibfield{author}{\bibinfo{person}{Shimin Chen} {and} \bibinfo{person}{Qin
  Jin}.} \bibinfo{year}{2015}\natexlab{}.
\newblock \showarticletitle{{Persistent B+-trees in Non-Volatile Main Memory}}.
\newblock \bibinfo{journal}{\emph{{PVLDB}}} \bibinfo{volume}{8},
  \bibinfo{number}{7} (\bibinfo{year}{2015}), \bibinfo{pages}{786--797}.
\newblock


\bibitem[\protect\citeauthoryear{Eisenman, Gardner, AbdelRahman, Axboe, Dong,
  Hazelwood, Petersen, Cidon, and Katti}{Eisenman et~al\mbox{.}}{2018}]%
        {eisenman2018reducing}
\bibfield{author}{\bibinfo{person}{Assaf Eisenman}, \bibinfo{person}{Darryl
  Gardner}, \bibinfo{person}{Islam AbdelRahman}, \bibinfo{person}{Jens Axboe},
  \bibinfo{person}{Siying Dong}, \bibinfo{person}{Kim Hazelwood},
  \bibinfo{person}{Chris Petersen}, \bibinfo{person}{Asaf Cidon}, {and}
  \bibinfo{person}{Sachin Katti}.} \bibinfo{year}{2018}\natexlab{}.
\newblock \showarticletitle{{Reducing DRAM Footprint with NVM in Facebook}}. In
  \bibinfo{booktitle}{\emph{Proceedings of the Thirteenth EuroSys Conference}}.
  ACM, \bibinfo{pages}{42}.
\newblock


\bibitem[\protect\citeauthoryear{Graefe, Guy, and Sauer}{Graefe
  et~al\mbox{.}}{2014a}]%
        {graefe2014instant}
\bibfield{author}{\bibinfo{person}{Goetz Graefe}, \bibinfo{person}{Wey Guy},
  {and} \bibinfo{person}{Caetano Sauer}.} \bibinfo{year}{2014}\natexlab{a}.
\newblock \showarticletitle{{Instant Recovery with Write-Ahead Logging: Page
  Repair, System Restart, and Media Restore}}.
\newblock \bibinfo{journal}{\emph{Synthesis Lectures on Data Management}}
  \bibinfo{volume}{6}, \bibinfo{number}{5} (\bibinfo{year}{2014}),
  \bibinfo{pages}{1--85}.
\newblock


\bibitem[\protect\citeauthoryear{Graefe and Kuno}{Graefe and Kuno}{2012}]%
        {DBLP:journals/pvldb/GraefeK12}
\bibfield{author}{\bibinfo{person}{Goetz Graefe} {and}
  \bibinfo{person}{Harumi~A. Kuno}.} \bibinfo{year}{2012}\natexlab{}.
\newblock \showarticletitle{{Definition, Detection, and Recovery of Single-Page
  Failures, a Fourth Class of Database Failures}}.
\newblock \bibinfo{journal}{\emph{{PVLDB}}} \bibinfo{volume}{5},
  \bibinfo{number}{7} (\bibinfo{year}{2012}), \bibinfo{pages}{646--655}.
\newblock


\bibitem[\protect\citeauthoryear{Graefe, Volos, Kimura, Kuno, Tucek,
  Lillibridge, and Veitch}{Graefe et~al\mbox{.}}{2014b}]%
        {graefe2014memory}
\bibfield{author}{\bibinfo{person}{Goetz Graefe}, \bibinfo{person}{Haris
  Volos}, \bibinfo{person}{Hideaki Kimura}, \bibinfo{person}{Harumi Kuno},
  \bibinfo{person}{Joseph Tucek}, \bibinfo{person}{Mark Lillibridge}, {and}
  \bibinfo{person}{Alistair Veitch}.} \bibinfo{year}{2014}\natexlab{b}.
\newblock \showarticletitle{{In-memory Performance for Big Data}}.
\newblock \bibinfo{journal}{\emph{{PVLDB}}} \bibinfo{volume}{8},
  \bibinfo{number}{1} (\bibinfo{year}{2014}), \bibinfo{pages}{37--48}.
\newblock


\bibitem[\protect\citeauthoryear{Lee, Lim, Song, Nam, and Noh}{Lee
  et~al\mbox{.}}{2017}]%
        {lee2017wort}
\bibfield{author}{\bibinfo{person}{Se~Kwon Lee}, \bibinfo{person}{K~Hyun Lim},
  \bibinfo{person}{Hyunsub Song}, \bibinfo{person}{Beomseok Nam}, {and}
  \bibinfo{person}{Sam~H Noh}.} \bibinfo{year}{2017}\natexlab{}.
\newblock \showarticletitle{{WORT: Write Optimal Radix Tree for Persistent
  Memory Storage Systems}}. In \bibinfo{booktitle}{\emph{15th USENIX Conference
  on File and Storage Technologies (FAST 17)}}. \bibinfo{pages}{257--270}.
\newblock


\bibitem[\protect\citeauthoryear{Leis, Haubenschild, Kemper, and Neumann}{Leis
  et~al\mbox{.}}{2018}]%
        {leis2018leanstore}
\bibfield{author}{\bibinfo{person}{Viktor Leis}, \bibinfo{person}{Michael
  Haubenschild}, \bibinfo{person}{Alfons Kemper}, {and} \bibinfo{person}{Thomas
  Neumann}.} \bibinfo{year}{2018}\natexlab{}.
\newblock \showarticletitle{{LeanStore: In-Memory Data Management Beyond Main
  Memory}}. In \bibinfo{booktitle}{\emph{2018 IEEE 34th International
  Conference on Data Engineering (ICDE)}}. IEEE, \bibinfo{pages}{185--196}.
\newblock


\bibitem[\protect\citeauthoryear{Lersch, Oukid, Schreter, and Lehner}{Lersch
  et~al\mbox{.}}{2017}]%
        {lersch2017rethinking}
\bibfield{author}{\bibinfo{person}{Lucas Lersch}, \bibinfo{person}{Ismail
  Oukid}, \bibinfo{person}{Ivan Schreter}, {and} \bibinfo{person}{Wolfgang
  Lehner}.} \bibinfo{year}{2017}\natexlab{}.
\newblock \showarticletitle{{Rethinking DRAM Caching for LSMs in an NVRAM
  Environment}}. In \bibinfo{booktitle}{\emph{European Conference on Advances
  in Databases and Information Systems}}. Springer, \bibinfo{pages}{326--340}.
\newblock


\bibitem[\protect\citeauthoryear{Mohan, Haderle, Lindsay, Pirahesh, and
  Schwarz}{Mohan et~al\mbox{.}}{1992}]%
        {mohan1992aries}
\bibfield{author}{\bibinfo{person}{C Mohan}, \bibinfo{person}{Don Haderle},
  \bibinfo{person}{Bruce Lindsay}, \bibinfo{person}{Hamid Pirahesh}, {and}
  \bibinfo{person}{Peter Schwarz}.} \bibinfo{year}{1992}\natexlab{}.
\newblock \showarticletitle{{ARIES: A Transaction Recovery Method Supporting
  Fine-Granularity Locking and Partial Rollbacks Using Write-Ahead Logging}}.
\newblock \bibinfo{journal}{\emph{ACM Transactions on Database Systems (TODS)}}
  \bibinfo{volume}{17}, \bibinfo{number}{1} (\bibinfo{year}{1992}),
  \bibinfo{pages}{94--162}.
\newblock


\bibitem[\protect\citeauthoryear{Ou, Chen, Xu, and H{\"a}rder}{Ou
  et~al\mbox{.}}{2014}]%
        {ou2014wear}
\bibfield{author}{\bibinfo{person}{Yi Ou}, \bibinfo{person}{Lei Chen},
  \bibinfo{person}{Jianliang Xu}, {and} \bibinfo{person}{Theo H{\"a}rder}.}
  \bibinfo{year}{2014}\natexlab{}.
\newblock \showarticletitle{Wear-Aware Algorithms for PCM-Based Database Buffer
  Pools}. In \bibinfo{booktitle}{\emph{International Conference on Web-Age
  Information Management}}. Springer, \bibinfo{pages}{165--176}.
\newblock


\bibitem[\protect\citeauthoryear{Oukid, Lasperas, Nica, Willhalm, and
  Lehner}{Oukid et~al\mbox{.}}{2016}]%
        {oukid2016fptree}
\bibfield{author}{\bibinfo{person}{Ismail Oukid}, \bibinfo{person}{Johan
  Lasperas}, \bibinfo{person}{Anisoara Nica}, \bibinfo{person}{Thomas
  Willhalm}, {and} \bibinfo{person}{Wolfgang Lehner}.}
  \bibinfo{year}{2016}\natexlab{}.
\newblock \showarticletitle{{FPTree: A Hybrid SCM-DRAM Persistent and
  Concurrent B-tree for Storage Class Memory}}. In
  \bibinfo{booktitle}{\emph{Proceedings of the 2016 International Conference on
  Management of Data}}. ACM, \bibinfo{pages}{371--386}.
\newblock


\bibitem[\protect\citeauthoryear{Pelley, Wenisch, Gold, and Bridge}{Pelley
  et~al\mbox{.}}{2013}]%
        {pelley2013storage}
\bibfield{author}{\bibinfo{person}{Steven Pelley}, \bibinfo{person}{Thomas~F
  Wenisch}, \bibinfo{person}{Brian~T Gold}, {and} \bibinfo{person}{Bill
  Bridge}.} \bibinfo{year}{2013}\natexlab{}.
\newblock \showarticletitle{{Storage Management in the NVRAM Era}}.
\newblock \bibinfo{journal}{\emph{{PVLDB}}} \bibinfo{volume}{7},
  \bibinfo{number}{2} (\bibinfo{year}{2013}), \bibinfo{pages}{121--132}.
\newblock


\bibitem[\protect\citeauthoryear{van Renen, Leis, Kemper, Neumann, Hashida, Oe,
  Doi, Harada, and Sato}{van Renen et~al\mbox{.}}{2018}]%
        {van2018managing}
\bibfield{author}{\bibinfo{person}{Alexander van Renen},
  \bibinfo{person}{Viktor Leis}, \bibinfo{person}{Alfons Kemper},
  \bibinfo{person}{Thomas Neumann}, \bibinfo{person}{Takushi Hashida},
  \bibinfo{person}{Kazuichi Oe}, \bibinfo{person}{Yoshiyasu Doi},
  \bibinfo{person}{Lilian Harada}, {and} \bibinfo{person}{Mitsuru Sato}.}
  \bibinfo{year}{2018}\natexlab{}.
\newblock \showarticletitle{{Managing Non-Volatile Memory in Database
  Systems}}. In \bibinfo{booktitle}{\emph{Proceedings of the 2018 International
  Conference on Management of Data}}. ACM, \bibinfo{pages}{1541--1555}.
\newblock


\bibitem[\protect\citeauthoryear{Wu, Jin, Yang, and Yue}{Wu
  et~al\mbox{.}}{2014}]%
        {wu2014app}
\bibfield{author}{\bibinfo{person}{Zhangling Wu}, \bibinfo{person}{Peiquan
  Jin}, \bibinfo{person}{Chengcheng Yang}, {and} \bibinfo{person}{Lihua Yue}.}
  \bibinfo{year}{2014}\natexlab{}.
\newblock \showarticletitle{{APP-LRU: A New Page Replacement Method for
  PCM/DRAM-Based Hybrid Memory Systems}}. In \bibinfo{booktitle}{\emph{IFIP
  International Conference on Network and Parallel Computing}}. Springer,
  \bibinfo{pages}{84--95}.
\newblock


\bibitem[\protect\citeauthoryear{Yang, Wei, Chen, Wang, Yong, and He}{Yang
  et~al\mbox{.}}{2015}]%
        {yang2015nv}
\bibfield{author}{\bibinfo{person}{Jun Yang}, \bibinfo{person}{Qingsong Wei},
  \bibinfo{person}{Cheng Chen}, \bibinfo{person}{Chundong Wang},
  \bibinfo{person}{Khai~Leong Yong}, {and} \bibinfo{person}{Bingsheng He}.}
  \bibinfo{year}{2015}\natexlab{}.
\newblock \showarticletitle{{NV-Tree: Reducing Consistency Cost for NVM-based
  Single Level Systems}}. In \bibinfo{booktitle}{\emph{13th USENIX Conference
  on File and Storage Technologies (FAST 15)}}. \bibinfo{pages}{167--181}.
\newblock


\end{thebibliography}

\end{document}